\documentclass[twocolumn,showpacs,10pt]{revtex4-1}%
\usepackage{amsmath,amssymb,graphicx}
\usepackage{amsmath}
\usepackage{amsfonts}
\usepackage{CJK}
\usepackage{amssymb}
\usepackage{graphicx}%
\providecommand{\U}[1]{\protect\rule{.1in}{.1in}}
\begin{document}

\title{Can a conventional optical camera realize turbulence-free imaging?}
\author{Deyang Duan and Yunjie Xia}
\email{yjxia@qfnu.edu.cn}
\affiliation{School of Physics and Physical Engineering, Qufu Normal University, Qufu 273165, China\\
Shandong Provincial Key Laboratory of Laser Polarization and Information
Technology, Research Institute of Laser, Qufu Normal University, Qufu 273165, China}
\begin{abstract}
Atmospheric turbulence is a serious problem for traditional optical imaging,
especially for satellite and aircraft-to-ground imaging. Here, we report
a novel and practical phenomenon in which turbulence-free images can be
reconstructed on a conventional optical camera based on the accumulation of
sunlight intensity and photon number fluctuation autocorrelation. Different
from conventional ghost imaging, this method can obtain turbulence-free
images, and its imaging speed is comparable to that of traditional optical
imaging. Moreover, by adding photon number fluctuation autocorrelation
algorithm software, almost all optical cameras, including mobile phone
cameras, can realize this function without changing the structure of the
original camera.

\end{abstract}
\maketitle

One of the most surprising consequences of quantum mechanics is the nonlocal
correlation of a multiparticle system observable in the joint detection of
distant particle detectors. Turbulence-free imaging is one such phenomena.
Turbulence-free imaging was first realized with a dual light path ghost
imaging framework [1,2]. Dual light path ghost imaging uses a correlation
measurement between two spatially correlated beams to reconstruct the ghost
image of an original object (Fig. 1a) [3,4]. One of the beams is called the
reference arm, which never illuminates the object and is directly measured by
a charged-coupled device (CCD). The other beam is the signal arm, which after
illuminating the object is measured by a detector without spatial resolution.
By a coincidence measurement of the signals from the two detectors, the ghost
image is restructured. Different from ghost imaging, traditional optical
imaging, which needs only one optical arm and one spatial resolution detector,
directly measures the intensity distribution of the light field (Fig. 1b).
Because of this, ghost imaging has some unique advantages that are important
for scientific research [5-7], medical imaging [8-10], remote sensing [11],
and night vision [12-14].

\begin{figure}[h]
\centering\includegraphics[width=1\linewidth]{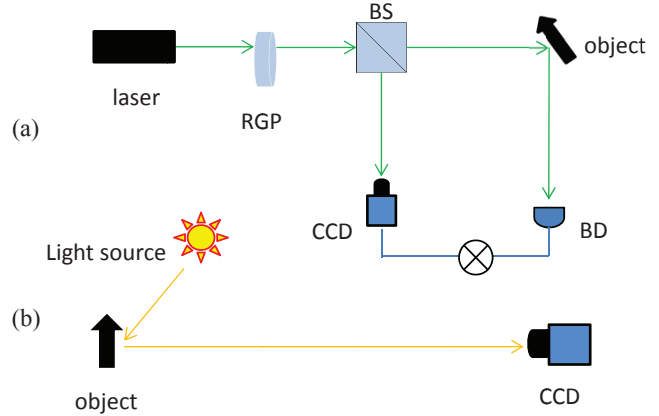}\caption{(a) Schematic
diagram of conventional ghost imaging; RGP: rotating ground plane, BS: beam
splitter, BD: bucket detector. (b) Schematic diagram of traditional optical
imaging. }%
\end{figure}

Although ghost imaging technology has many advantages, there is no property
comparable to the incredible turbulence-free imaging property [1,2,15]. This
practical property is an important milestone for optical imaging because any
fluctuation index disturbance introduced in the optical path will not affect
the image quality. However, conventional optical imaging cannot overcome the
influence of atmospheric turbulence without image processing technology (see
Appendix I for details). Previous works show that turbulence-free ghost
imaging requires certain conditions [16,17]. For example, turbulence-free
imaging requires that the two optical paths need to go through the same
turbulence. Consequently, turbulence-free images cannot be obtained by
conventional computational ghost imaging or single-pixel imaging [18]. Another
condition requires that the transverse coherence size of the illumination
light be less than that of the turbulence. Fortunately, the coherence size of
sunlight is sufficiently small. For example, the coherence size of sunlight
with a central wavelength of 550 nm on the ground is 0.08 mm. Consequently,
sunlight is an ideal light source for turbulence-free imaging. Previous works
show that ghost imaging can be realized by measuring the accumulation of the
fluctuation intensity of sunlight at a certain time [17,19],
which makes this problem solveable by ghost imaging.

The intensity correlation of the light field is a critical resource to realize
turbulence-free imaging. However, conventional optical cameras cannot directly
measure the intensity correlation of the light field [20]. Consequently,
traditional optical imaging was affected by interference of atmospheric
turbulence until the introduction of image processing. Now, satellite remote
sensing images (e.g., Google Earth: 104.96E, 26.59N) still have distortions
caused by atmospheric turbulence. Can turbulence-free imaging be achieved with
only a conventional cameras and sunlight? In this article, we give an
affirmative answer to this challenging question; i.e., a turbulence-free image
can be obtained by processing the data collected by a conventional optical
camera with a photon number fluctuation autocorrelation algorithm.
Furthermore, we show that almost all conventional optical cameras, including
mobile phones, can realize turbulence-free imaging by only adding this
software and without changing their structures.

\begin{figure}[h]
\centering\includegraphics[width=1\linewidth]{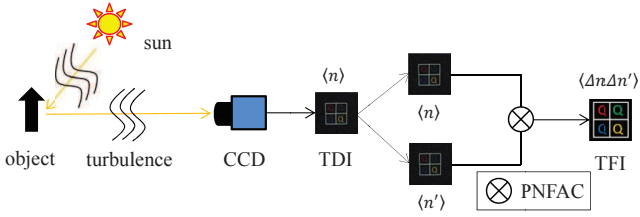}\caption{Setup of
turbulence-free imaging based on a conventional optical camera. TDI:
turbulence distorted image, TFI: turbulence-free image, PNFAC: photon number
fluctuation autocorrelation. }%
\end{figure}

The conceptual arrangement of our method based on a conventional optical
camera and sunlight is illustrated in Fig. 2. A beam of sunlight illuminates
an object, and then its reflected light is directly received by a conventional
optical CCD camera. Generally, we assume that atmospheric turbulence exists in
the light path between the sun and the object and between the object and the
camera. According to the imaging principle of optical cameras, the light field
received by a camera can be expressed as
\begin{equation}
E\left(  x\right)  =\int\int dx_{i}dx_{o}E_{i}\left(  x_{i}\right)
g_{o}e^{\phi_{1}}T\left(  x_{o}\right)  g_{i}e^{\phi_{2}},
\end{equation}
where $x_{o}$ and $x_{i}$ are the transverse coordinates in the object and
image planes, respectively. $g_{o}$ is the Green's function that propagates
the light from the light source (coordinate $x_{i}$) to point $x_{o}$ on the
object plane. $T(x)$ is the object. $g_{i}$ is the Green's function that
propagates the light field from $x_{o}$ to $x_{i}$ on the camera plane.
$e^{\phi_{1}}$ and $e^{\phi_{2}}$ represent the atmospheric turbulence between
the sun and the object and between the object and the camera, respectively
[13-19,21,22]. The data are processed by the photon number fluctuation
autocorrelation algorithm, and the reconstructed image can be expressed as%

\begin{equation}
G^{\left(  2\right)  }=\left\langle \Delta n\Delta n^{^{\prime}}\right\rangle
=\frac{1}{m}\left\vert \sum_{m}E_{m}^{\ast}\left(  x\right)  E_{m}^{^{\prime}%
}\left(  x\right)  \right\vert ^{2},
\end{equation}
where $m$ represents the number of measurements and $\Delta n$ and $\Delta
n^{^{\prime}}$ represent the photon number fluctuations of $E$ and $E^{\prime
}$, respectively. Equation 2 shows that a turbulence-free image can be
obtained by our method (see Appendix II for details).

The photon number fluctuation autocorrelation algorithm is briefly described
below (see Appendix III for details). The software first calculates the
average counting numbers per each short time window $\overline{n}$. Two
virtual logic circuits (post-neg identifiers) classify the counting numbers
per window as positive and negative fluctuations based on $\overline{n}$.
Thus, we have%

\begin{align}
\Delta n_{\alpha}^{(+)} &  =\left\{
\begin{array}
[c]{c}%
n_{\alpha}-\overline{n},if,n_{\alpha}>\overline{n}\\
0,otherwise
\end{array}
\right.  \nonumber\\
\Delta n_{\alpha}^{(-)} &  =\left\{
\begin{array}
[c]{c}%
n_{\alpha}-\overline{n},if,n_{\alpha}<\overline{n}\\
0,otherwise,
\end{array}
\right.
\end{align}
where $\alpha=1$ to $m$ and is used to label the $\alpha$th short time window.
$m$ is the total number of time windows. Then,
we define the following quantities for the statistical correlation
calculations of%
\begin{align}
\left(  \Delta n\Delta n^{\prime}\right)  _{\alpha}^{\left(  ++\right)  } &
=\left\vert \Delta n_{\alpha}^{(+)}\Delta n_{\alpha}^{^{\prime}(+)}\right\vert
,\nonumber\\
\left(  \Delta n\Delta n^{\prime}\right)  _{\alpha}^{\left(  --\right)  } &
=\left\vert \Delta n_{\alpha}^{(-)}\Delta n_{\alpha}^{^{\prime}(-)}\right\vert
,\nonumber\\
\left(  \Delta n\Delta n^{^{\prime}}\right)  _{\alpha}^{\left(  +-\right)  }
&  =\left\vert \left(  \overline{n}-\Delta n_{\alpha}^{(+)}\right)  \left(
\overline{n}-\Delta n_{\alpha}^{^{\prime}(-)}\right)  \right\vert ,\nonumber\\
\left(  \Delta n\Delta n^{^{\prime}}\right)  _{\alpha}^{\left(  -+\right)  }
&  =\left\vert \left(  \overline{n}-\Delta n_{\alpha}^{(-)}\right)  \left(
\overline{n}-\Delta n_{\alpha}^{^{\prime}(+)}\right)  \right\vert .
\end{align}
Thus, the corresponding statistical average of $\left\langle \Delta n\Delta
n^{^{\prime}}\right\rangle $ is%

\begin{align}
\left\langle \Delta n\Delta n^{^{\prime}}\right\rangle  &  =\frac{1}{m}\left[
\sum_{\alpha=1}^{m}\left(  \Delta n\Delta n^{^{\prime}}\right)  _{\alpha
}^{\left(  ++\right)  }+\sum_{\alpha=1}^{m}\left(  \Delta n\Delta n^{^{\prime
}}\right)  _{\alpha}^{\left(  --\right)  }\right. \nonumber\\
&  +\left.  \sum_{\alpha=1}^{m}\left(  \Delta n\Delta n^{^{\prime}}\right)
_{\alpha}^{\left(  +-\right)  }+\sum_{\alpha=1}^{m}\left(  \Delta n\Delta
n^{^{\prime}}\right)  _{\alpha}^{\left(  -+\right)  }\right]  .
\end{align}

A schematic of the experimental setup is shown in Fig. 2. It is a typical and
traditional optical imaging schematic, except for the addition of data
processing software. In this experiment, atmospheric turbulence is introduced
by adding heating elements underneath any or all optical paths, as illustrated
in Fig. 2. Heating of the air causes temporal and spatial fluctuations in the
index of refraction and makes the traditional image of the object jitter about
randomly on the image plane causing the image to become distorted. The
sunlight illuminates the object (the letter ``Q'') located at $x_{o}$. The
photons reflected from the object are collected and counted by a conventional
industrial camera (The Imaging Source DFK23U618). The path from the object to
the detector over heating elements is 0.4 m. The CCD camera is controlled by
software to collect data. A photon number fluctuation autocorrelation
algorithm program is used to process the data.

Fig. 3 shows a set of typical experimental results. Fig. 3A(a) and Fig. 3B(a)
show two traditional optical images without turbulence, i.e., $\left\langle
n\right\rangle $. Figs. 3A(b-e) show four distorted traditional images caused
by atmospheric turbulence. Correspondingly, Figs. 3B(b-e) show four
turbulence-free images reconstructed by the photon number fluctuation
autocorrelation algorithm; i.e., $\left\langle \Delta n\Delta n^{^{\prime}%
}\right\rangle $. To quantitatively analyse the quality of the reconstructed
image, the structural similarity index (SSIM) is used as an evaluation index.
Table 3C shows that the image reconstructed by this method can indeed be
regarded as a turbulence-free image.

\begin{figure}[h]
\centering\includegraphics[width=1\linewidth]{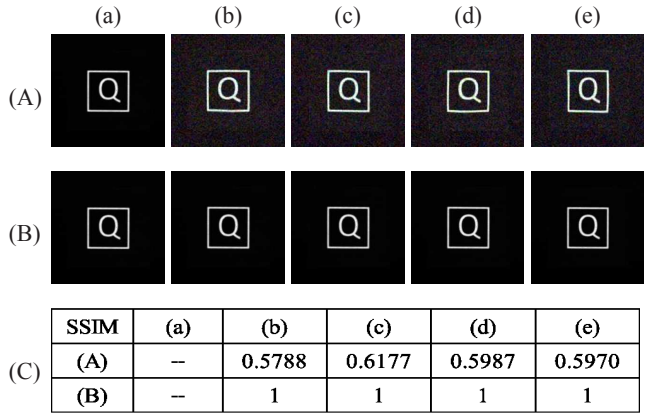}\caption{(A-a) and
(B-a) show two traditional optical images without atmospheric turbulence.
A(b-e) are four traditional optical images distorted by atmospheric
turbulence. B(b-e) correspond to four images reconstructed by 100
measurements. (C) The SSIM of traditional optical images distorted by
atmospheric turbulence and the turbulence-free images obtained by photon
number fluctuation autocorrelation. }%
\end{figure}\begin{figure}[h]
\centering\includegraphics[width=1\linewidth]{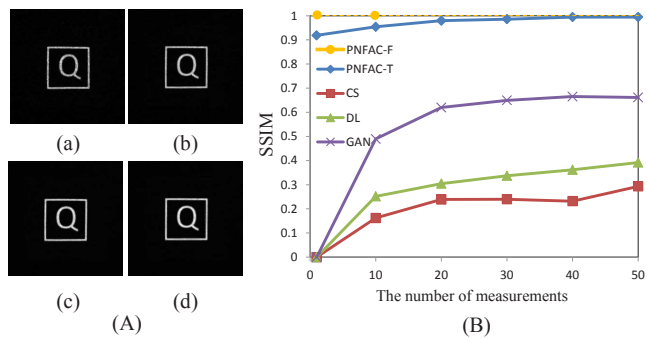}\caption{(A): Four
turbulence-free images reconstructed from 1, 10, 20, and 30 measurements in
the atmospheric turbulence environment. (B): The SSIM curves of the images
reconstructed by this method without atmospheric turbulence (PNFAC-F) and with
atmospheric turbulence (PNFAC-T). The SSIM curves of the reconstructed ghost
images of compressed sensing (CS), deep learning (DL), and generative
adversarial networks (GANs) with different measurements in the environment
without atmospheric turbulence. }%
\end{figure}\begin{figure}[h]
\centering\includegraphics[width=0.9\linewidth]{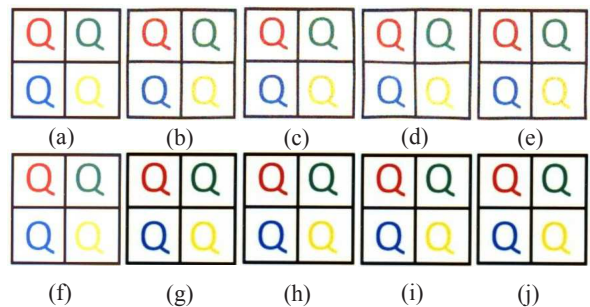}\caption{The color
turbulence-free images reconstructed by our method. (a) and (f) are two
traditional colour images without atmospheric turbulence. (b)-(e) are four
traditional colour image distorted by atmospheric turbulence. (f)-(j) are the
corresponding turbulence-free images reconstructed by the photon number
fluctuation autocorrelation algorithm. (g)-(j) are images reconstructed from
50 measurements. }%
\end{figure}\begin{figure}[h]
\centering\includegraphics[width=0.9\linewidth]{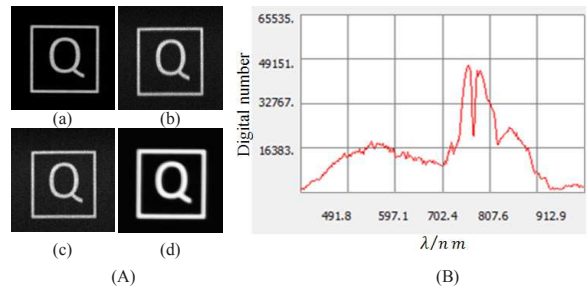}\caption{(A): (a) A
turbulence-free image obtained by a digital single lens reflex camera (Nikon,
D750). (b) A turbulence-free hyperspectral image obtained by a hyperspectral
camera (Dualix Spectral Imaging, GaiaField). The centre wavelength is 532.4
nm. (c) A turbulence-free night vision image obtained by a conventional
infrared camera (Intevac, NIR). The transmission band of the long wave pass
filter is 800 nm-2500 nm (Daheng Optics GCC-300123). (d) A turbulence-free
image obtained by a conventional mobile phone (Huawei, Mate 30 Pro). These
four images were reconstructed from 20 measurements. (B): The solar spectrum
of the experimental environment. }%
\end{figure}

Due to the novel imaging mode, another significant advantage of this method is
that it has a fast imaging speed and can meet the requirements of practical
applications. The imaging speed of this method in a turbulent environment is
almost comparable to that of traditional optical imaging in an environment
without turbulence. The imaging speed of this method in turbulent environments
is significantly better than that of conventional ghost imaging with
optimization schemes in environments without turbulence [23-29]. Fig. 4A
presents a set of experimental results. The experimental results show that a
turbulence-free image with quality comparable to that of traditional optical
imaging can be obtained with only 1 measurement, which is an almost impossible
challenge for ghost imaging frameworks. The results of several other typical
ghost imaging methods without turbulence and this method are presented and
compared in Fig. 4B. Surprisingly, the image obtained by this method is
completely consistent with the traditional optical image after 1 measurement
in the environment without atmospheric turbulence. Moreover, a traditional
optical imaging video and a video obtained with this method are shown and
compared in the appendix. Fig. 5 is a set of colour turbulence-free images
reconstructed by this method. The results show that this method can be
directly applied to existing colour optical cameras.
Another traditional optical imaging video and a video obtained with this
method are shown in the appendix.

To demonstrate the feasibility of this method on a variety of cameras, we
selected several typical cameras for experiments. The results are shown in
Fig. 6. Here, we present the black-and-white image results because some
cameras, such as infrared cameras, can only directly output black-and-white
images. Fig. 6 shows that even a conventional mobile phone camera can realize
turbulence-free imaging.
\begin{figure}[h]
\centering\includegraphics[width=0.9\linewidth]{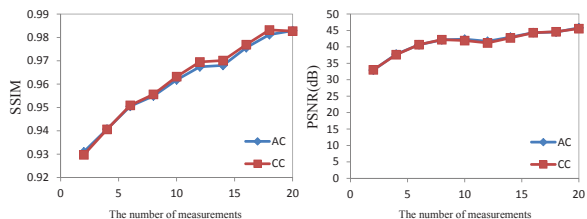}\caption{The SSIM and PSNR
curves of the experimental results with photon number fluctuation autocorrelation
algorithm (AC) and photon number fluctuation crosscorrelation algorithm (CC).
}%
\end{figure}

It should be emphasized that the photon number fluctuation autocorrelation algorithm
can be used when the camera collects only one data. However, the results obtained by
the photon number fluctuation autocorrelation algorithm and the photon number fluctuation
crosscorrelation algorithm are consistent when the camera collects two or more
data. See Appendix III for theoretical demonstration. To quantitatively demonstrate this property, we use the structural similarity (SSIM)
and PSNR to measure the experimental results (Fig. 7).

These experiments represent the first demonstration of turbulence-free imaging
realized on a conventional optical camera. Using the accumulation of intensity
fluctuations of the light field at a certain time and the photon number
fluctuation autocorrelation algorithm, our results illustrate how a
conventional optical camera without any structural changes can produce a
turbulence-free colour image. Moreover, we demonstrate that the imaging speed
of this method is the same as that of traditional optical imaging in
environments without atmospheric turbulence and is comparable to that of
traditional optical imaging in environments with atmospheric turbulence. This
imaging speed meets the practical application requirements of existing
cameras. This technique provides a promising solution to images affected by
atmospheric turbulence. More importantly, this technique can be applied quickly.

This project was supported by the National Natural Science Foundation (China)
under grant nos. 11704221, 11574178 and 61675115 and the Taishan Scholar
Project of Shandong Province (China) under grant no. tsqn201812059.

The authors declare that there are no conflicts of interest related to this article.

\end{document}